\documentclass[12pt]{article}
\usepackage{amsfonts,amsmath}
\usepackage{amssymb}
\usepackage[hypertex] {hyperref}

\makeatletter \@addtoreset{equation}{section} \makeatother

\def\theequation{\thesection.\arabic{equation}}

\newcommand{\be}{\begin{equation}}
\newcommand{\ee}{\end{equation}}
\newcommand{\bee}{\begin{eqnarray}}
\newcommand{\beee}{\begin{array}}
\newcommand{\eee}{\end{eqnarray}}
\newcommand{\eeee}{\end{array}}

\newcommand{\bV}{{V}}

\newcommand{\un}{{\underline{n}}}

\newcommand{\ga}{\alpha}
\newcommand{\pa}{{\dot{\ga}}}

\newcommand{\gb}{\beta}

\newcommand{\W}{{\cal W}}
\newcommand{\K}{{\cal K}}
\newcommand{\F}{{\cal F}}

\newcommand{\ie}{{\it i.e.,} }
\newcommand{\ls}{\!\!\!\!\!\!}

\newcommand{\gvep}{\varepsilon}

\newcommand{\gs}{\sigma}

\newcommand{\go}{\omega}

\newcommand{\q}{\,,\qquad}

\newcommand{\dga}{{\dot{\alpha}}}

\newcommand{\nn}{\nonumber}

\newcommand{\half}{\frac{1}{2}}

\newcommand{\p}{\partial}
\newcommand{\pp}{\partial^*_Z{}}

\newcommand{\f}{\frac}
\newcommand{\B}{{\cal B}}%%

\newcommand{\wick}{{normal\,\,}}

\newcommand{\U}{\Upsilon}
\newcommand{\ups}{\upsilon}
\newcommand{\bu}{\bar{\upsilon}}

\newcommand{\dr}{{\rm d}}

\newcommand{\Sp}{{\mathcal H}}
\newcommand{\Spl}{{\mathcal H}^{loc}}

\begin{document}

\begin{flushright}
%\texttt{Started on May 14}
%\\
{\small FIAN/TD/17-14}
\end{flushright}
\vspace{1.7 cm}

\begin{center}
{\large\bf Star-Product Functions in  Higher-Spin
Theory
and Locality}

\vspace{1 cm}

{\bf  M.A.~Vasiliev}\\
\vspace{0.5 cm}
{\it
 I.E. Tamm Department of Theoretical Physics, Lebedev Physical Institute,\\
Leninsky prospect 53, 119991, Moscow, Russia}

%\vspace{0.6 cm}
%didenko@lpi.ru, matveev@lpi.ru, vasiliev@lpi.ru \\
\end{center}

\vspace {1cm}

{\it $\phantom{MMMMMMMMMMMMMMMMMMMMMMMMMMMM}$ To my mother}

\vspace{1.2cm}

\vspace{0.4 cm}

\begin{abstract}
Properties of the functional classes  of star-product elements
associated with higher-spin gauge fields and gauge parameters  are elaborated.
Cohomological interpretation  of the nonlinear higher-spin equations is given.
An algebra $\Sp$, where solutions of the nonlinear higher-spin equations are valued,
is found. A conjecture on the classes of star-product functions underlying
(non)local maps and gauge transformations in the nonlinear higher-spin theory is proposed.

\end{abstract}

\newpage
\tableofcontents

\newpage

\section{Introduction}
\label{intro}

In this paper we identify functional classes  of star-product elements relevant
to the analysis of nonlinear higher-spin (HS) equations.
Our results give the cohomological interpretation of the
terms responsible for interactions and shed light on the important issue of (non)locality
 in HS theories leading to a conjecture on the classes of (non)local functionals,
 field redefinitions and gauge transformations. This identification also restricts
 possible generalizations of the HS equations.

Importance of the issue of locality in HS theory was realized in
\cite{Prokushkin:1998bq} where it was shown that, by a seemingly
local field redefinition induced by the so-called integrating flow
found in the same paper for  the $3d$ HS theory, it is possible to
get rid of currents from the {\it r.h.s.} of HS field equations
including the stress tensor in the  spin-two sector. It was argued
in \cite{Prokushkin:1998bq} that  this result implies that the
field transformation induced by the integrating flow is nonlocal. In
\cite{Prokushkin:1999xq}, where this issue was further elaborated in
terms of currents, such transformations were called {\it
pseudolocal}. In the $AdS$ background they have a form of an
infinite derivative expansion \be \label{phi'} \phi\to\phi'=\phi
+\sum_n a_{nm} (\rho D)^n \phi\,(\rho D)^m \phi+\ldots\,, \ee where
$\rho$ is the $AdS$ radius and  $D$ is the space-time covariant
derivative. The problem is to find restrictions on the coefficients
$a_{nm}$ distinguishing between truly non-local and generalized local
field redefinitions which may contain an infinite number of terms
but the coefficients $a_{nm}$
 decrease fast enough with $n$ and $m$. Note that the problems in $AdS_d$ and
 Minkowski space are essentially different  since the expansion (\ref{phi'})
does not make sense in the naive limit $\rho\to \infty$.

In the unfolded form of the  HS theories  suggested in
\cite{more,Prokushkin:1998bq,Vasiliev:2003ev} the space-time dependence is encoded
in  additional twistor-like variables $Z^A$ and $Y^A$ where the meaning of the
index $A$ depends on the model. In the twistor-like variables,  (\ref{phi'}) is
substituted by
\be
\label{phi''}
\phi\to\phi'=\phi +\sum_{nmkl} b_{nmkl} \Big (\Big(\f{\p}{\p Z}\Big)^n
\Big( \f{\p}{\p Y}\Big)^m \phi\Big) \Big (\Big (\f{\p}{\p Z}\Big)^k \Big (\f{\p}{\p Y}\Big)^l
\phi\Big )+\ldots\,
\ee
and the problem is to find  restrictions on the coefficients
$b_{nmkl}$. Being reformulated in terms of  appropriate classes of functions,
this is one of the goals of this paper. Specifically, we identify such restrictions
on the maps between star-product elements in HS theory, that are algebraically consistent
(form an algebra) and rule out the nonlocal transformations resulting from the integrating
flow of \cite{Prokushkin:1998bq} and similar. These will be conjectured to represent local
maps in the HS theory. The class of allowed gauge transformations is also identified.

The rest of the paper is organized as follows.
In Section \ref{HS star product} we recall properties of the HS star product  underlying
nonlinear HS theories.   In Section \ref{Nonlinear Higher-Spin Equations} the structure
of the field equations for the nonlinear $AdS_4$ HS theory  is recalled
and their cohomological interpretation is discussed.
In Section \ref{functspa}  functional spaces of star-product
elements  underlying our construction are introduced. The algebra $\Sp$ where HS fields
are valued is defined in Section \ref{Sp}.
The locality conjecture is formulated in Section \ref{locality}. Conclusions and
perspectives are discussed in Section \ref{conc}.
Details of the proof of {\it Lemma 5} are presented in Appendix A.
A relation between the Weyl star product and the algebra $\Sp$ is briefly discussed in
Appendix B.

\section{HS star product}
\label{HS star product}

HS equations were formulated in \cite{more,Prokushkin:1998bq,Vasiliev:2003ev} in terms of
the associative \emph{HS star product}  $*$ which acts on functions of two
variables $Z_A$ and $Y_A$
\be
\label{star2}
(f*g)(Z;Y)=\frac{1}{(2\pi)^{M}}
\int d^{M} U\,d^{M} V \exp{[iU^A V^B C_{AB}]}\, f(Z+U;Y+U)
g(Z-V;Y+V) \,,
\ee
where $C_{AB}=-C_{BA}$ ($A,B,\ldots = 1,2,\ldots,M$)
is  nondegenerate  allowing to raise and lower indices
\be
Y^A = C^{AB}Y_B\q Y_A= Y^B C_{BA}\,,
\ee
and
$ U^A $, $ V^B $ are real integration variables. Star product (\ref{star2}),
normalized so that $1$ is its unit element, \ie $f*1 = 1*f =f$,
 yields a specific realization of the Weyl algebra\footnote{Weyl algebra is the
 algebra of oscillators. It should not be confused with the Weyl star product
 describing the product law in a specific frame of the Weyl algebra.}
\be
\label{zzyy}
[Y_A,Y_B]_*=-[Z_A,Z_B ]_*=2iC_{AB}\,,\qquad
[Y_A,Z_B]_*=0\q[a,b]_*=a*b-b*a\,
\ee
and possesses a supertrace operation
\be
\label{str}
str (f(Z;Y)) = \frac{1}{(2\pi)^{M}}
\int d^{M} U\,d^{M} V \exp{[-iU^A V^B C_{AB}]}\, f(U;V) \,
\ee
obeying the cyclic property
\be
\label{cycl}
str(f*g ) = str (g*f)
\ee
provided that the
coefficients of the expansions of $f(Z;Y)$ and $g(Z;Y)$
in powers of $Z_A$ and $Y_A$ are (anti)commuting for  $f(Z;Y)$ and $g(Z;Y)$
(odd)even under $f(-Z;-Y)= (-1)^{\pi_f}f(Z;Y)$.

An important property of  star product (\ref{star2}) is that it
admits the inner Klein operator
\be
\label{ups}
\Upsilon = \exp i Z_A Y^A \,,
\ee
which obeys
\renewcommand{\U}{\Upsilon}
\be
\label{UU}
\U *\U =1,
\ee
\be
\label{[UF]}
\U *f(Z;Y)=f(-Z;-Y)*\U\,.
\ee

The Klein operator $\U_Y$ for the
 star product of $Z$-independent functions, which amounts to the Weyl star product
\be
\label{starweyl}
(f*g)(Y)=\frac{1}{(2\pi)^{M}}
\int d^{M} U\,d^{M} V \exp{[iU^A V^B C_{AB}]}\, f(Y+U) g(Y+V)\,,
\ee
is the $\delta$-function \cite{BerezShub}
\be
\U_Y =(2\pi)^{\f{M}{2}} \delta^M(Y)\,.
\ee
Indeed, from (\ref{starweyl}) it follows that
\be
\delta^M (Y)* f (Y) =f(-Y)*\delta^M (Y)\,.
\ee
$\U_Y$ squares to  unity
\be
\label{}
\U_Y*\U_Y = 1.
\ee
(For $\hbar\neq 1$ reinserted into the definition of the star product,
the {\it r.h.s.} of this relation is proportional to
$\hbar^{-M}$ and, as anticipated,  becomes infinite in the classical limit.)

An important property of the Klein operator is that it  generates the
Fourier transform:
\be
f(Y)*\U_Y = \tilde f (Y):=
\frac{1}{(2\pi)^{\f{M}{2}}}
\int d^{M} U\, \exp{[-iU_A Y^A]}\, f(U)  \,.
\ee

For $Z$-independent elements $f(Z;Y)=f(Y)$ (\ref{str}) gives the well-known result
\cite{V3}
\be
\label{strweyl}
str (f(Y))=f(0)\,.
\ee
 From here it follows that
though $\U_Y$ is well behaving with respect to the star product its supertrace is divergent
\be
str(\U_Y)= \infty\sim \delta^M(0)\,.
\ee

Analogously, one can define
\be
\U_Z =(2\pi)^{\f{M}{2}} \delta^M(Z)\,.
\ee
Klein operator (\ref{ups}) results from the star product (\ref{star2})
of the Klein operators in the $Y$- and $Z$-sectors
\be
\U = \U_Y*\U_Z\,.
 \ee

From (\ref{strweyl}) it follows that in the Weyl star product $\star$, which is the direct
product of the Weyl star products in the $Y$ and $Z$ sectors,
\be
str(\U_Y\star \U_Z)=str(\U_Y)str(\U_Z)\sim\delta^{2M}(0)\,.
\ee
On the other hand, the supertrace is insensitive to a  basis of the Weyl algebra, \ie to
the form of the star product. Hence $str(\U)$ remains divergent as $\delta^{2M}(0)$ in
the HS star product. Analogously $str (f)=\infty$ for any  $f(Z;Y)$  behaving
in $Y$ and $Z$ like $\U$.
This fact plays the key r\'ole in \cite{funk} where
 the supertrace of nontrivial invariant functionals is demanded to be divergent.

\section{HS equations in $AdS_4$}
\label{Nonlinear Higher-Spin Equations}

\subsection{Nonlinear system and its cohomological  interpretation}

 HS theory in $AdS_4$  was formulated
in  \cite{more} in terms of the zero-form $ B(Z;Y;\K|x)$,
space-time connection one-form $W(Z;Y;\K|x)$ and  connection one-form $S(Z;Y;\K|x) $
in the $Z$-space. $W(Z;Y;\K|x)$ and $S(Z;Y;\K|x) $ can be combined into
the total connection one-form
\be
\W = \dr_x+ \theta^\un W_\un (Z;Y;\K|x)+ \theta^A S_A (Z;Y;\K|x)\q \dr_x =
\theta^\un \f{\p}{\p x^\un}\,,
\ee
where all differentials $dZ^A\equiv \theta^A$ and $dx^\un\equiv \theta^\un$
are anticommuting. In this section,
$A=1,\ldots ,4$ and $\un =0,\ldots 3$ are indices of $4d$ Majorana spinors and
vectors, respectively. $\K=~\!\!(k,\bar{k})$ denotes a pair of Klein operators
that reflect chiral spinor indices of every
$U^A=(u^\ga,\bar u^\dga)$ with $U^A=(Y^A,Z^A,\theta^A)$,
 $u^\ga= (y^\ga, z^\ga, \theta^\ga )$, $\bar u^\pa =
(\bar y^\pa, \bar z^\pa, \bar \theta^\pa )$,
\bee
\label{kk}
k* u^\ga = -u^\ga* k\,,\quad &&
k *\bar u^\pa = \bar u^\pa *k\,,\quad
\bar k *u^\ga = u^\ga *\bar k\,,\quad
\bar k *\bar u^\pa = -\bar u^\pa *\bar k\,,\nn\\
&&\quad k*k=\bar k*\bar k = 1\,,\quad
k*\bar k = \bar k *k\,.
\eee
Note that relations (\ref{kk})
provide the definition of the star product with $k$ and $\bar k$.

The simplest version of the nonlinear HS equations of \cite{more} is
\be
\label{SS}
\W*\W= \F(\B)\q \F(\B)=-i \big ( \theta_A \theta^A + \eta
\delta^2(\theta_z)  \B* k*\ups + \bar \eta
\delta^2(\bar \theta_{\bar z}) \bar \B *\bar k *\bu \big)
\,,
\ee
\be
\label{SB}
\W*\B=\B*\W\,,
\ee
where
\be
\delta^2(\theta_z)= \half \theta_\ga \theta^\ga\q
\delta^2(\bar \theta_{\bar z})= \half\bar \theta_\dga \bar\theta^\dga\,,
\ee
$\eta=\exp [i\varphi]$, $\varphi\in [0,\pi)$ (the absolute value of $\eta
$ as well as a factor of $-1$ can be absorbed
into a redefinition of $\B$) leads to a class of pairwise nonequivalent nonlinear HS
theories. The left and right inner Klein operators
\be
\label{kk4}
\ups =\exp i z_\ga y^\ga\,,\qquad
\bu =\exp i \bar{z}_\dga \bar{y}^\dga\,
\ee
commute with $\theta^A$ and obey
\be
\label{[uf]}
\ups *f(z,\bar{z};y,\bar{y})=f(-z,\bar{z};-y,\bar{y})*\ups\,,\quad
\bu *f(z,\bar{z};y,\bar{y})=f(z,-\bar{z};y,-\bar{y})*\bu\,,
\ee
\be
\ups *\ups =\bu *\bu =1\q \ups *\bu = \bu*\ups\,.
\ee

Equations (\ref{SS}), (\ref{SB}) can be extended to
$\W(\theta;Z;Y;\K|x)$ and $\B(\theta;Z;Y;\K|x)$ being
differential forms of arbitrary odd and even total degrees, respectively
(both in $\theta_Z$ and in $\theta_x$).
(Such an extension was considered, {\it e.g.}, in \cite{333,Boulanger:2011dd}.)
In this case  equations (\ref{SS}), (\ref{SB}) are invariant under the following
gauge transformations
\be
\label{dw}
\delta \W = [\W\,, \gvep]_*\,+  \xi^N \f{\p \F(\B)}{\p \B^N}\q\delta \B = \{\W\,,\xi\}_* + [\B\,, \gvep]_* \,,
\ee
where
$\gvep(\theta;Z;Y;\K|x)$ and $\xi(\theta;Z;Y;\K|x)$, which are even
and odd functions of $\theta$, respectively, are gauge parameters
associated with $\W(\theta;Z;Y;\K|x)$ and $\B(\theta;Z;Y;\K|x)$.
$N$ is the infinite multiindex running over all components of $\B$.

To clarify the origin of the Klein operators in the   HS
equations  we rewrite
Eq.~(\ref{SS}) in the form
\be
\label{SSco}
\W*\W= -i (\theta_A \theta^A + \delta^2(\theta_z)\delta^2(z)*  \phi +
\delta^2(\bar \theta_{\bar z}) \delta^2(\bar z)* \bar \phi )
\,,
\ee
where $\phi$ and $\bar\phi$  commute with $\W$
up to $\theta_z$-- and $\bar \theta_{\bar z}$--dependent terms, respectively,
 that do not affect  the compatibility conditions
of (\ref{SSco}) since $\theta_z^3=\bar\theta_{\bar z}^3=0$. The {\it r.h.s.} of (\ref{SSco})
admits simple cohomological interpretation within the perturbative analysis of the
HS equations.

Indeed, consider the standard vacuum solution with $\B=0$ and
\be
\label{vac1}
\W_0 = \dr_x + Q +  W_0(Y|x)\,,
\ee
where
\be
\label{Q}
Q:=\theta^A Z_A
\ee
and the space-time one-form $W_0(Y|x)$ (the differentials $\theta_x$ are
implicit) is some solution to the flatness equation
\be
\label{W0eq}
\dr_x W_0(Y|x) +W_0(Y|x)*W_0(Y|x) =0\,.
\ee
By Eq.~(\ref{zzyy}), the star-commutator with
$Q$ is proportional to the de Rham derivative
in  $Z^A$
\be
\label{dzf}
Q * f(Z;Y) -(-1)^{deg_f} f(Z;Y)*Q = -2i \dr_Z f(Z;Y)\q
\dr_Z = \theta^A \f{\p}{\p Z^A}\,,
\ee
where $deg_f$ is the form degree of $f$.
We observe that $\delta^2(\theta_z)\delta^2(z)$ in (\ref{SSco})
describes the de Rham cohomology of
$\dr_z$.\footnote{This is a particular case of
the de Rham cohomology associated with any submanifold $M'\subset M$
described by the
equation $F_{M'}(X)=0$ where $X$ are local coordinates of $M$. Namely,
$V_{M'}\delta(F_{M'}(X))$, where $V_{M'}$ is the volume form on
$T^*(M)/T^*(M')$, represents
the de Rham cohomology of $M$.} Indeed, $\delta^2(\theta_z)\delta^2(z)$ is
$\dr_Z$-closed since $\theta^3_z=0$ and is not $\dr_Z$-exact since
the $\delta^2(z)$ cannot be integrated in  well-behaving functions.
This means that the interaction terms on the {\it r.h.s.} of Eq.~(\ref{SS})
form a consistent but nontrivial source.
Moreover, for the Weyl-Moyal star product, equation (\ref{SS}) admits no
meaningful solution at all. Star product
(\ref{star2}), that mixes $Z$ and $Y$ variables in a nontrivial way, makes  system
(\ref{SS})  solvable.
The conjecture of Section \ref{locality} suggests however that the interaction
terms still cannot be removed by a local field redefinition.
Note that relevance of a cohomological interpretation of the HS equations
 was pointed out long ago in \cite{Vasiliev:1990bu}.

\subsection{Perturbative analysis}
\label{sketch}
Let
\be
\label{W1}
\W = \W_0 +\W_1 +\ldots\q \B =\B_1 +\ldots\,,
\ee
where $\W_1$ and $\B_1$ are first-order fluctuations.
Linearized  equations (\ref{SS}), (\ref{SB}) are
\be
\label{dW1}
{\rm d} \W_1 + W_0 *\W_1 +\W_1*W_0 =
-i \Big (\eta \delta^2(\theta_z)  \B_1* k*\ups +
\bar \eta \delta^2(\bar \theta_{\bar z})  \B_1* \bar k *\bu \Big )\,,
\ee
\be
\label{dB1}
{\rm d} \B_1 +W_0 *\B_1 -\B_1 * W_0 =0\q \dr=\dr_Z+\dr_x\,.
\ee

The one-form sector of (\ref{dB1}) yields
\be
\label{BC}
\B_1^0(Z;Y;\K|x) = C^0(Y;\K|x)\,
\ee
and
\be
\label{dc}
{\rm d} C^0 (Y;\K|x)+W_0(Y|x) *C^0(Y;\K|x) - C^0(Y;\K|x)* W_0(Y|x) =0\,,
\ee
where $\B_1^0(Z;Y;\K|x)$ and  $C^0(Y;\K|x)$ are zero-forms.

The main tool for the perturbative analysis  is provided by the
standard homotopy formula
\be \label{homm1} \dr_Z g(\theta_Z; Z; Y)=f( \theta_Z; Z; Y)
\quad \Longrightarrow\quad
g(\theta_Z; Z; Y)= \pp f  +\dr_Z\gvep + g(0;0;Y)\,,
\ee
where
\be
\label{pzp}
\pp f :=\dr^*_Z H(f)\q H(f):=
\int_0^1 dt t^{-1} f(t \theta_Z; tZ; Y)\q
 \dr_Z^* = Z^A\frac{\partial}{\partial \theta^{A}}
\,.
 \ee
The term  $\dr_Z\gvep $ in Eq.~(\ref{homm1})
describes the freedom in exact forms while $g(0;0;Y)$ represents
the  de Rham cohomology.
 Eq.~(\ref{homm1}) is valid provided that the homotopy integral over $t$ converges,
which, in accordance with the Poincar\'e lemma, is true if  $g(0;0;Y)=0$. Note that
$\dr_Z^*\dr_Z^*=0 $ implies
\be
\label{pzpz}
 \pp \pp =0\,.
\ee

The two-form sector of (\ref{dW1}) yields for the one-form $\W^1_1 =\W_1^{1,0} +\W_1^{0,1}$,
\be
\label{w110}
\W_1^{1,0} = \half \eta \int_0^1 d\tau \,\tau e^{i\tau z_\ga y^\ga}z_\ga
\theta^\ga
C(-\tau z,\bar y;\K)k + \half \bar \eta \int_0^1 d\bar \tau \,\bar \tau \exp^{i\bar \tau\bar z_\dga \bar
y^\dga}\bar z_\dga \bar \theta^\dga
C(y,-\bar \tau \bar z;\K)\bar k\,,
\ee
\be
\label{w101}
\W_1^{0,1} =\go^1 -\f{i}{2} \pp \{W_0\,,\W_1^{1,0}\}\,,
\ee
where the one-form $\go^1(\theta_x;Y;\K|x)$ describes physical HS gauge fields
valued in  the cohomology of $\dr_Z$
(the term with $\dr_x$ in (\ref{dW1}) does not contribute to
the second term in (\ref{w101}) due to (\ref{pzpz})).
Plugging (\ref{w101}) into the $\theta_x^2$ sector of (\ref{dW1}) yields
the so-called First On-Shell Theorem  which
imposes dynamical equations on the spin $s>1$ frame-like fields contained in
$\go^1(\theta_x;Y;\K|x)$ (for more detail see  \cite{Vasiliev:1999ba} and references therein).

The aim of this section is to recall  two general features.

The first is that dynamically nontrivial (physical) components of the HS
fields like $C^0(Y;\K|x)$ and $\go^1(\theta_x;Y;\K|x)$ are in the $\dr_Z$ cohomology.
Other components of the HS
fields are expressed via the physical components by the HS equations.
Indeed,
since $\dr$ contains ${\rm d_Z}$ (\ref{dzf}),  equations (\ref{dW1}) and (\ref{dB1})
express all components of $\W_1$ and $\B_1$ that are not ${\rm d_Z}$ closed via
other fields. ${\rm d_Z}$-exact fields are pure gauge with respect to the gauge
transformations (\ref{dw}). Hence, the remaining {\it physical} fields are in  the
${\rm d_Z}$-cohomology.
By Poincar\'e Lemma, physical fields are independent of both $Z^A$ and
$\theta^A$, \ie
\be
C(Y;\K|x):= \B_1(Z;Y;\K|x)\Big |_{\theta_Z=Z=0}\q
\go(\theta_x;Y;\K|x):= \W_1(\theta;Z;Y;\K|x)\Big |_{\theta_Z=Z=0}\,.
\ee
(Note that this consideration  ignores the issue controlled by the so-called
$\gs_-$--cohomology that the HS equations  impose further
constraints expressing some of the physical
fields via space-time derivatives of the other.)

The second is that the perturbative solution of the
HS equations leads to homotopy integrals over $\tau$ in formulae like (\ref{w110}). Higher perturbations  lead to multiple homotopy integrals which
are in the core of  the analysis of star-product functional classes in  the sequel.

\section{Functional spaces}
\label{functspa}
\subsection{Algebra $\bV_{0,0}$}
\label{H+}
Functions of the variables $Y$ and $Z$, emerging in the perturbative analysis of HS equations,
 are nonpolynomial. Indeed, application of (\ref{homm1}) to terms containing the
Klein operators (\ref{kk4}) gives rise to functions of the form
\be
\label{ff}
f(Z;Y)=\int_0^1 d \tau \varphi( Z;Y;\tau)
\exp{i\tau Z_A Y^A}\,,
\ee
where indices $A,B,\ldots = 1,2,\ldots , M$
can take any even number of values
(e.g., $Z_A$, $Y_A$ can  denote each of the pairs of spinors $z_\ga$, $y_\ga$ or
$\bar z_\dga$, $\bar y_\dga$) and
\be
\varphi (Z;Y;\tau)= \sum_{n,m=0}^\infty \varphi_{A_1,\ldots, A_n\,,
B_1,\ldots ,B_m}(\tau)
Z^{A_1}\ldots Z^{A_n} Y^{B_1}\ldots Y^{B_m}\,
\ee
is a polynomial or power series in $Z$, $Y$ with
the coefficients $\varphi_{A_1,\ldots ,A_n\,, B_1,\ldots ,B_m}(\tau)$
 integrable in $\tau$. Distributions in $\tau$ are also allowed.

 As  observed originally  in \cite{Pr},   elements (\ref{ff}) form a closed
algebra under the HS star product (\ref{star2}), \ie their star products  are
free of divergencies and belong to the class (\ref{ff}).
This is specific for the HS star product (\ref{star2}) and may not be true for
other star products associated with different ordering prescriptions.
Indeed, a possible divergency of the star product of Gaussian
exponentials is due to a potential degeneracy of the Gaussian bilinear form
in the integration variables $U$ and $V$ in (\ref{star2}). However,
because of the  form of  star product (\ref{star2}),
the $\tau$-dependent exponential in (\ref{ff}) does not contribute to
the quadratic part in $U$ and $V$  since  $U_A U^A=V_A V^A=0$.
For other star products, which typically involve $4M$ integration variables
in the analogue of (\ref{star2}) (cf. formula (\ref{starw}) for the Weyl star product),
 this mechanism does not work and divergencies can appear.

An elementary computation yields
\be
\label{fstarf}
(f_1 * f_2 )(Z;Y) =
\int d\tau_{1,2} \varphi_{1,2}( Z;Y;\tau_{1,2})
\exp{i\tau_{1,2} Z_A Y^A}\,,
\ee
where
\bee
\label{vphiphi}
\varphi_{1,2}( Z;Y;\tau_{1,2})=
 &&\ls\f{1}{(2\pi)^M}\int d{\tau_1} d{\tau_2} dSdT \delta(\tau_{1,2} - \tau_1\diamond \tau_2)
 \exp i S_A T^A\nn\\
&&\ls\ls\ls\varphi_1((1-\tau_2) Z - \tau_2 Y + S;
(1-\tau_2) Y -\tau_2 Z +S;\tau_1)
\nn\\
&&\ls\ls\ls
\varphi_2((1-\tau_1) Z
+ \tau_1 Y -T;
\tau_1 Z + (1-\tau_1) Y  +T;\tau_2)\,,
\eee
and
\be
\label{circ}
a\diamond b =a+b -2ab=a(1-b) +b(1-a)\,
\ee
can be interpreted as a product in $\mathbb{R}$ or $\mathbb{C}$.
It is  obviously commutative
\be
a\diamond b=b\diamond a\,
\ee
 and associative, being inherited from the associative star product.
Setting
\be
\label{ga} a = \half(1-\ga)\q b = \half(1-\gb) \
\ee
 (\ref{circ}) yields the usual product $\ga\gb$
with $a=0,\half$ and $1$  mapped, respectively, to $\ga=1,$
$0$, and $-1$. This complies with the facts  that, with respect to the star
product, $1$ is the
unit element, $\exp{i {Z}_A {Y}^A}$ is the involutive Klein
operator (\ref{ups}) and
\be \label{Fo} F=\exp{\f{i}{2}{Z}_A {Y}^A}
\ee
is the Fock vacuum obeying
\be  (Y_A +Z_A)* F =0\q F*
(Y_A -Z_A)=0\q F* \exp{(ia {Z}_A {Y}^A)} = \exp{(i a {Z}_A {Y}^A})*
F = F\,.
\ee

{}Since the segment $[-1,1]$ is invariant under
multiplication, it follows that
\be
\label{prodpr}
a, b \in [0,1]\qquad\Longrightarrow \qquad a\diamond b \in [0,1]\,
\ee
as is also obvious from
\be
\label{eq}
1-a\diamond b = (1-a)(1-b) +ab\,.
\ee

For $\varphi_1(Z;Y;\tau)$ and $\varphi_2(Z;Y;\tau)$
polynomial in $Z$ and $Y$, that are integrable in $\tau$,
$\varphi_{1,2}$ is also a polynomial integrable in $\tau_{1,2}$.
Hence functions (\ref{ff}) form an algebra called $\Sp^{tot}$.

Consider the subspace $\bV_{0,0}\subset \Sp^{tot}$ of functions of the form
\be
\label{ffreg}
f(Z;Y)=\int_0^1 d\tau \phi( \tau Z;(1-\tau)Y;\tau)
\exp{i\tau Z_A Y^A}\,
\ee
 with
$\phi( W; U;\tau)$  regular in $W$ and $U$ and integrable in $\tau$.
Being accompanied by the factor of $\tau$ and $1-\tau$, the
dependence on $Z$ and $Y$ on the {\it r.h.s.} of (\ref{ffreg})  trivializes
at $\tau \to 0$ and $\tau\to 1$, respectively. Such a behavior turns out to be most
appropriate for the perturbative analysis of the HS theory.

The functions $\phi(W,U,\tau)$
 localized at $\tau=0$ or $\tau=1$ are of special interest.
Functions (\ref{ffreg}) with $\phi(W;U;\tau)$ proportional
to $\delta(\tau)$ yield $Z$--independent  elements
\be
\label{fb}
f(Z;Y) =  g(Y)\,,
\ee
while those proportional to $\delta(1-\tau)$ have the form
\be
\label{fe}
f(Z;Y) =h(Z)\,\exp{iZ_A Y^A} \,.
\ee

The following remarkable fact is true:\\

\noindent
{\it Theorem 1}:
$\bV_{0,0}$ forms an associative algebra with respect to the star product (\ref{star2}).
\phantom{{\it Theorem 1}:\quad}Elements with $\phi( W; U;\tau)$ polynomial in $W, U$ form its subalgebra.

\noindent
For the proof we observe that (\ref{vphiphi}) yields\bee
\label{vphiphireg}
\varphi_{1,2}( W;U;\tau_{1,2})=
 &&\ls\f{1}{(2\pi)^M}\int dSdT\exp  i S_A T^A \int_0^1 d\tau_1 d\tau_2 \delta( \tau_{1,2} - \tau_1\diamond \tau_2)
 \nn\\
&&\ls\ls\ls\ls\phi_1(\tau_1[(1-\tau_2) W - \tau_2 U + S];
(1-\tau_1)[
(1-\tau_2) U-\tau_2 W   +S];\tau_1)
\nn\\
&&\ls\ls\ls\ls\ls\phi_2(\tau_2[(1-\tau_1) W + \tau_1 U -T];
(1-\tau_2)[\tau_1 W + (1-\tau_1) U  +T];\tau_2)\,.
\eee
Elementary inequalities following from
Eqs.~(\ref{circ}) and (\ref{eq})
\be
\label{ineq1}
\tau_1\diamond \tau_2\geq (1-\tau_1)\tau_2\geq 0\q
\tau_1\diamond \tau_2 \geq (1-\tau_2)\tau_1\geq 0\,,
\ee
\be
\label{ineq2}
1-\tau_1\diamond \tau_2 \geq (1-\tau_1)(1-\tau_2)\geq 0\q
1-\tau_1\diamond \tau_2\geq \tau_1 \tau_2\geq 0\,
\ee
imply that
\be
\label{0-}
(1-\tau_1)\tau_2 = \alpha(\tau_1,\tau_2) \tau_1\diamond \tau_2 \q
(1-\tau_2)\tau_1 = \beta(\tau_1,\tau_2) \tau_1\diamond \tau_2\,,
\ee
\be
\label{1-}
(1-\tau_1)(1-\tau_2) = \gamma(\tau_1,\tau_2) (1- \tau_1\diamond \tau_2 ) \q
\tau_1\tau_2 = \rho(\tau_1,\tau_2) (1-\tau_1\diamond \tau_2)\,
\ee
with
\be
\label{abgr}
\alpha(\tau_1,\tau_2),\, \beta(\tau_1,\tau_2),\, \gamma(\tau_1,\tau_2),\,
\rho(\tau_1,\tau_2) \in [0,1]\,.
\ee
This implies that, for any $f_1$ and $f_2$ of the form (\ref{ffreg}),
$f_1 * f_2$ also has the form (\ref{ffreg}) with
\bee
\phi_{1,2}( W;U;\tau)=&&\ls \f{1}{(2\pi)^M}
\int dSdT\exp iS_A T^A \int_0^1 d\tau_1 d\tau_2
\delta( \tau - \tau_1\diamond \tau_2)
\nn\\
&&\ls\ls\ls\ls
\phi_1(\beta W - \rho U +\tau_1 S;
\gamma U -\alpha W  +(1-\tau_1)S;\tau_1)
\nn\\
&&\ls\ls\ls\ls\ls
\phi_2(\alpha W + \rho U -\tau_2 T;
\beta W + \gamma U  +(1-\tau_2)T;\tau_2)\,
\eee
provided that
\be
\int dSdT\exp iS_A T^A
\phi_1(W  +a S;
U +(1-a)S;\tau_1)
\phi_2(V  -b T;
Y + (1-b)T;\tau_2)\,\q a,b\in [0,1]\,
\ee
is well defined (converges).
Due to (\ref{abgr}),
$
\phi_{1,2} (W,U,\tau )
$
is  integrable in $\tau$. Hence, $f_1 *f_2\in \bV_{0,0}$.
Clearly,
$
\phi_{1,2} (W,U,\tau )
$
 is polynomial in $W$ and $U$ if
$\phi_1(W;U;\tau)$ and  $\phi_2(W;U;\tau)$ were  $\Box$\\

The property  inherited from the Klein operator is that
not every  element of $\bV_{0,0}$ admits  supertrace.
Indeed, by virtue of (\ref{str}),
\be
str (f)=\f{1}{(2\pi)^{M}}\int d^{M} Z d^{M} Y \int_0^1 d\tau \phi( \tau Z;(1-\tau)Y;\tau)
\exp{[-i(1-\tau) Z_A Y^A}]\,.
\ee
 Changing the integration variables
$
Y^A = (1-\tau )^{-1} U^A$, $Z^A = V^A
$
yields
\be
\label{strhs}
str (f)=\f{1}{(2\pi)^{M}}\int d^{M} U d^{M} V \int_0^1 d\tau (1-\tau)^{-M}
\phi( \tau V;U;\tau)
\exp{-i( V_A U^A)}\,.
\ee
Since $\phi (W;U;\tau)$ is regular in its arguments, $str (f)$ is
well defined provided that the integral over $\tau$ converges at $\tau=1$,
\ie $\phi(  V;U;\tau)$  appropriately tends to zero
at $\tau \to 1$.
In Section \ref{Sp} it is shown that, generally, only the
logarithmic divergency matters, and the algebra $\Spl_0$
 free of trace divergencies is introduced.

Let us stress that,
beyond the algebra $\bV_{0,0}$, the degree of divergency
of the supertrace may depend on the degree of $Y$ (\ie  spin) while in $\bV_{0,0}$
this does not happen just
because the $Y$-dependence   in (\ref{ffreg}) has been rescaled by  $1-\tau$.

\subsection{Spaces $\bV_{k,l}$}
\label{spaces}

It is useful to consider  the space $\bV_{k,l}$ of such star-product elements
(\ref{ffreg}) that $\phi(W;U;\tau)$ scales as $\tau^k$ at $\tau\to 0$ and  $(1-\tau)^l$
at $\tau\to 1 $. More precisely, we allow (poly)logarithmic dependence
on $\tau$ and $1-\tau$ at $\tau\to 0$ and $\tau\to 1$, respectively, with the
convention that it does not affect the indices $k$ and $l$.
We  consider $\bV_{k,l}$ with both  positive and negative $k$ and/or $l$
assuming however  that the function $\varphi(Z;Y;\tau)$ in
(\ref{ff}) is integrable in $\tau$.
For example, for the spaces
$\bV_{-1,0}$ and $\bV_{0,-1}$ this is the case if $\phi(0;U;\tau)=0$
and $\phi(W;0;\tau)=0$, respectively.

Thus, $\bV_{k,l}$ contains such $f$ (\ref{ffreg}) that
$\tau^{-k} (1-\tau)^{-l} \phi(W;U;\tau)$ has non-negative scalings at $\tau\to 0,1$.
The spaces $\bV_{k,\,l}$ with non-integer indices $k$ and $l$
are also allowed. With this definition
\be
\label{vep}
\bV_{k+\epsilon,\,l} \subset \bV_{k,\,l} \q
\bV_{k,\,l+\epsilon} \subset \bV_{k,\,l}  \qquad \forall
\epsilon >0\,.
\ee

Via  the decomposition  $1=\tau +(1-\tau)$ any $f\in \bV_{k,\,l}$ can be represented in the form
\be
\label{decomb}
f\in \bV_{k,\,l}:\qquad f=g+h\q g\in \bV_{k+1,\,l}\q h\in  \bV_{k,\,l+1}\,.
\ee
Repeated application of this formula along with (\ref{vep}) gives\\

\noindent
{\it Lemma 1}:
\be
\bV_{k,l} =\bV_{k,\infty} \cup \bV_{\infty,l}\,,
\ee
where
\be
\bV_{k,\infty}\subset \bV_{k,l} \quad \forall l\q
\bV_{\infty,l} \subset \bV_{k,l}  \quad \forall k\,,
\ee
\ie
elements of $\bV_{k,\infty}$ ($\bV_{\infty,l}$) have $\phi(W;U;\tau)$ that
tend to zero at $\tau\to 1$ ($\tau\to 0$) faster than any power
of $1-\tau$ ($\tau)$.\\

Alternatively, {\it Lemma 1} follows from the decomposition
\be
\int_0^1 d\tau\phi(\tau) = \int_{-\infty}^\infty
d\tau \vartheta(\tau)\vartheta(1-\tau)\phi(\tau)\,,
\ee
\be
\label{deco}
\vartheta(\tau)\vartheta(1-\tau) \phi(W;U;\tau) =\vartheta(\tau)\vartheta(a-\tau)\phi(W;U;\tau) +
\vartheta(\tau-a)
\vartheta(1-\tau)
\phi(W;U;\tau)\q \forall a\in (0,1)\,,
\ee
where $\vartheta(\tau)$ is the step function. Indeed, here the first and second terms
are  identically
zero in some neighborhood of $\tau=1$ and $\tau=0$, respectively.

For distributions  we assign
\be
\phi(W;U;\tau)\sim \delta(\tau):\quad f(Z;Y)\in \bV_{-1, \infty}
 \,,
\ee
\be
\phi(W;U;\tau)\sim \delta(1-\tau):\quad f(Z;Y)\in \bV_{\infty, -1}\,.
\ee
The derivatives $\delta^k(\tau)$ and $\delta^k(1-\tau)$ of $\delta(\tau)$
and $\delta(1-\tau)$ are assigned  to $\bV_{-1-k,\,\infty}$
and $\bV_{\infty\,,-1-k,}$, respectively.

The spaces $ \bV_{k,l}$ have the fundamental composition property  expressed by\\

\noindent
{\it Theorem 2:}
\be
\label{VV}
\bV_{k_1,l_1} * \bV_{k_2,l_2}
\subset \bV_{\min(k_1, l_2)+\min(k_2,l_1)+1\,,
\,\min(k_1,k_2)+\min(l_1,l_2 )+1}\,.
\ee
\\
This follows from  formula (\ref{vphiphireg}) and
 inequalities (\ref{ineq1}), (\ref{ineq2}) along with
 the simple fact that the integral
$$
 \int_0^1 d\tau_1 \int_0^1 d\tau_2 \delta( \tau - \tau_1\diamond \tau_2)=-\log((1- 2\tau)^2)
$$
behaves as $\tau$ at $\tau\to 0$ and $1-\tau$ at $\tau\to 1$ $\Box$\\

Formula (\ref{VV}) has  useful consequences. In particular
\be
\bV_{k,l} * \bV_{k,l} \subset \bV_{2\min(k, l)+1
\,,k+l+1}\,.
\ee
For $\bV_{k,l}$ with $k\geq -1$, $l\geq -1$ from here it follows using (\ref{vep}) that
\be
\label{V*V}
\bV_{k,l} * \bV_{k,l} \subset \bV_{\min(k, l)
\,,\max (k,l)}\q k\geq -1\,,\,\, l\geq -1\,.
\ee
Hence,  $\bV_{k,l}$ with $l\geq k\geq -1$ forms an algebra
\be
\label{vkl}
\bV_{k,l}*\bV_{k,l}\subset \bV_{k,l}\q l\geq k\geq -1\,.
\ee
The fact proven in Section \ref{H+} that $\bV_{0,0}$ is an
algebra is a particular case of (\ref{vkl}).

{}From (\ref{vep}), (\ref{VV}) and (\ref{V*V}) it  follows that
\be
\bV_{l,k}*\bV_{k,l}\subset \bV_{l,k}\q
\bV_{k,l}*\bV_{l,k}\subset \bV_{l,k}\q
\bV_{l,k}*\bV_{l,k}\subset \bV_{k,l}\q l\geq k\geq -1\,,
\ee
implying along with (\ref{vkl}) that
\be
\label{hlk}
W_{l,k}=
W_{k,l} := \bV_{k,l} \cup \bV_{l,k}\q k\geq -1\,,\,\,l\geq -1\,
\ee
is an algebra. Interesting algebras of this type are
$W_{-1,0}$ and $W_{-1,\infty}$.

{}For $f$ (\ref{ffreg}), formula (\ref{vphiphireg}) with $f_2=\U$ yields
\be
f*\U =\int_0^1 d \tau \phi\big (-(1-\tau)Y; -\tau Z;1-\tau)\exp i\tau Z_A Y^A\,.
\ee
Along with (\ref{[UF]}) this has an important consequence
\be
\label{l0}
\U* \bV_{k,l}= \bV_{k,l}*\U= \bV_{l,k}\,.
\ee
Together with  (\ref{VV}) this gives\\

\noindent {\it Lemma 2}:
\be
\label{L31}
\bV_{-1,\infty} * \bV_{k,l} = \bV_{k,l} * \bV_{-1,\infty} = \bV_{k,l}\,,
\ee
\be
\bV_{\infty,-1} * \bV_{k,l} = \bV_{k,l} * \bV_{\infty,-1} = \bV_{l,k}\,.
\ee
\\

{}From (\ref{L31}) it follows in particular that the
star product of a function that depends
only on $Y$ with an element of $\bV_{k,l}$ belongs to $\bV_{k,l}$.

\subsection{The derivative and homotopy}
\label{difhom}
Now we extend  consideration to differential forms in the
$Z$ space.
Let $f(\theta_Z;Z;Y)\in \bV_{k,\,l,p}$ if $f(\theta_Z;Z;Y)$ is a $p$-form
with coefficients in $\bV_{k,\,l}$.

An important property of  $Q$ (\ref{Q}) is\\

\noindent
{\it Lemma 3}: $Q\in\bV_{-2,\,\infty,1}$.
\\
\noindent
This is because
\be
Q=\int_0^1 d\tau \delta (\tau) \tau^{-1} (\tau \theta^A Z_A)\,,
\ee
where both $\delta(\tau)$ and  $\tau^{-1}$
bring negative contribution to the first index of $\bV_{-2,\,l,1}$.
\\

Since the graded  star-commutator $[Q\,,\ldots ]_*$   is equivalent
to  $\dr_Z=\theta^A \f{\p}{\p Z^A}$,
we arrive at\\

\noindent
{\it Lemma 4}:
$
[Q \,,\bV_{k,l,p}]_* \subset \bV_{k+1,l-1,p+1}\,.
$
\\
Here   $k$ increases because the $Z$-differentiation of
$f(Z;Y)$ (\ref{ffreg}) brings one power of $\tau$ while $l$
decreases because the $Z$-differentiation of the exponential
in (\ref{ffreg}) brings one power of $ Y$  requiring a factor of
$1-\tau$ $\Box$
\\

Suppose that $f\in \bV_{k,l,p}$  is $Q$-closed. By  homotopy formula (\ref{pzp}),
a solution to equation (\ref{homm1})
$
\dr_Z g = f\,\
$
for $f$ (\ref{ffreg}) is
\be
\label{pfdz}
\pp f(\theta_Z;Z;Y)=
Z^A\f{\p}{\p \theta^A}\int_0^1 \f{ds}{s} \int_0^1 d\tau \phi(s \theta_Z;  s \tau
Z;(1-\tau)Y;\tau)
\exp i{s \tau Z_A Y^A}\,.
\ee
An elementary analysis sketched in Appendix A proves\\

\noindent
{\it Lemma 5}:
\be
\label{hommap}
\pp \bV_{k,l,p} \subset \bV_{ min(p-1,k)-1\,,\,l+1\,,\,p-1}\,.
\ee
\label{Lemma 6}

\subsection{Inner and boundary spaces}
\label{inbou}

To distinguish between $\phi(W;U;\tau)$ localized at the boundary of the
segment $\tau\in [0,1]$ and those smooth at $\tau =0$ or
$\tau = 1$, the functions $\phi(W;U;\tau)$  should be further specified. Factors of
$W_A U^A$ in $\phi(W;U;\tau)$ can be removed by the partial
integration over $\tau$ in (\ref{ffreg})
 implying that
\be
\label{equiv}
\exp{(i\tau Z_A Y^A)}\left [
i \vartheta(\tau) \vartheta (1-\tau) Z_A Y^A \chi(\tau Z;(1-\tau) Y;\tau) +
\f{\p}{\p \tau} \big (\vartheta(\tau) \vartheta (1-\tau)
\chi(\tau Z;(1-\tau) Y;\tau) \big )\right ] \sim 0\,.
\ee
As a result, the representation (\ref{ffreg}) can  be achieved
with such $\phi(W;U;\tau)$ that its {\it inner part} $\phi^{in}(W;U;\tau)$,
containing an additional factor of $\tau(1-\tau)$
 to cancel the denominator in
 \be
 Z_A Y^A = \tau^{-1} (1-\tau)^{-1} W_A U^A\q W_A=\tau Z_A\,,\quad U_A=(1-\tau) Y_A\,,
 \ee
obeys the condition
\be
\label{cla}
\f{\p^2}{\p W_A \p U^A} \phi^{in}(W;U;\tau)=0\,.
\ee

 The spaces  of $f\in \bV_{k,l}$ (\ref{ffreg}) with smooth  functions
$\phi (W;U;\tau)$  obeying (\ref{cla}) and those localized at $\tau=0 $
or $1$
will be called {\it inner space} $\widetilde V_{k,l}$ and {\it boundary space}
$\overline V_{k,l}$, respectively. They can play a r\'ole in  the analysis of
HS dynamics as discussed  in Section \ref{conc}.
Taking into account the dependence on $\tau$,
condition (\ref{cla}) restricts  $\phi(W;U;\tau)$
to functions of as many variables as the original unrestricted function $f(Z;Y)$,
\ie $\widetilde V_{k,l}$ is as large as the space of functions of $Z,Y$. On the other
hand, $\overline V_{k,l}$ is the space of functions that depend either only on
$Y$ or only on $Z$ since the dependence on $Z$ and $Y$ in (\ref{ffreg}) trivializes at
$\tau=0$ and $\tau=1$, respectively ({\it cf}. Eqs.~(\ref{fb}), (\ref{fe})).

In particular, {\it physical} fields
defined in  Section \ref{sketch} as belonging to $\dr_Z$-cohomology
obey\\

\noindent
{\it Lemma 6:} Physical fields belong to $\overline\Sp^{phys}:=\overline \bV_{-1,\infty,0}$.
$\overline\Sp^{phys}$ is an algebra.

\section{HS field algebra $\Sp$ and local algebra $\Spl$}

\label{Sp}

The case with a single homotopy parameter $\tau$ in (\ref{ffreg}) considered
so far  is not  most general  since, like in
HS equations (\ref{SS}), there may be several Klein operators associated with
different de Rham cohomologies in the $Z$-space. In that case we introduce several homotopy
parameters $\vec{\tau}= (\tau_1 , \tau_2\ldots )$ with the respective
multiindices $\vec{k}= (k_1,k_2,\ldots) $, $\vec{l}=
(l_1,l_2,\ldots )$ and $\vec{p}= (p_1,p_2,\ldots) $ of $\bV_{\vec{k},
\vec{l},\vec{p}}$. If the homotopy parameters appear in the
combination  $\exp{i\tau Z_A Y^A}$, for the example of $4d$ HS model we  set
\be
\int_0^1 d \tau \varphi( Z;Y;\tau)
\exp{i\tau Z_A Y^A}= \int_0^1 d \tau_1 \int_0^1 d \tau_2
\varphi( Z;Y;\tau_1)\delta(\tau_1 -\tau_2)\exp i[{\tau_1 z_\ga y^\ga
+\tau_2 \bar z_\dga \bar y^\dga }]\,,
\ee
assigning the indices $k_1$, $k_2$ and $l_1$, $l_2$  freely
at the condition that $k_1+k_2 = k$ and $l_1+l_2 = l$ where $k$ and $l$ control
the behavior in $\tau$.
With these multi-index notations our analysis applies to general HS systems
with $M\to \vec{M}=(M_1\,,M_2,\ldots)$\,, $M=M_1+M_2 +\ldots $\,.
For the sake of  simplicity   we will use the single-index notation in the sequel.

\subsection{HS field algebra}
\label{FA}
The results of Sections \ref{spaces} and \ref{difhom} allow us
to identify the space  of fields $\Sp$
appropriate for the perturbative analysis of the HS equations:
\be
\label{spn}
\Sp :=\oplus_{p=0}^M \Sp_p\q \Sp_p :=\bV_{p-1,M-p-1,p}\,.
\ee

Using that any  $p$-form in
$\theta_Z$ with $p>M$ is zero, from  formula (\ref{VV}) at $p+q\leq M$
follows \\

\noindent
{\it Lemma 7}:
$
\Sp_p *\Sp_q \subset \Sp_{p+q}\,.
$
\\

{\it Lemma 7} respects the $\mathbb{Z}$-grading of the exterior
algebra and  implies the important\\

\noindent
{\it Theorem 3}:
$\Sp$ is an algebra, \ie
$
\Sp*\Sp \subset \Sp\,.
$
\\

$\Sp$ will be called \emph{HS field algebra}.  Note that by {\it Lemma 6}
physical fields belong to $\overline\Sp^{phys}\subset\Sp_0\subset\Sp$.

{\it Lemmas 2} and {\it 3} imply\\

\noindent
{\it Lemma 8}:
$
[Q \,,\Sp_{p}]_* \subset \Sp_{p+1}\q
\pp \Sp_{p} \subset \Sp_{p-1}\q \pp \Sp_{0}=0\,
$\\

which has a consequence\\

\noindent
{\it Theorem 4}: $\Sp$  is invariant under the action of
the homotopy operator $\pp$ and  derivative $\dr_Z$.\\

One of the important conclusions of this paper is that the operator $Q\in
\bV_{-2,\forall l,1}$ does
not belong to $\Sp$. {\it Theorem 4} then implies
that $Q$ induces an outer derivation of $\Sp$.
Correspondingly, the HS connection $\W$ should be written in the form
\be
\label{W'}
\W = \dr_x +Q +\W'\q \W'\in \Sp\,.
\ee

The central result of this section is\\

\noindent
{\it Theorem 5}: The fields $\W'$ and $\B$  resulting from the perturbative
solution of the HS equation with the homotopy operator (\ref{pzp}) belong to $\Sp$
in all orders of the perturbative expansion.

The proof  follows from {\it Theorems 3}, {\it 4} along with the fact
that, since
$\delta^M(\theta_Z) \Upsilon \in \Sp_M\subset \Sp$, by {\it Theorem 3}  the term
$\delta^M(\theta_Z)\B*k* \Upsilon$ on the {\it r.h.s.} of the HS equations
belongs to $\Sp$. Also one should take into account that the physical fields
belong to $\Sp$ by {\it Lemma 6} $\Box$
\\

Another important consequence of {\it Theorems 3} and {\it 4}  is\\

\noindent
{\it Theorem 6}: Gauge transformations (\ref{dw}) with $\varepsilon,\xi \in \Sp$ leave
the HS fields  in $\Sp$.\\

A distinguishing property of the homotopy operator $\pp$ %ERR
is that it maps  the spaces $\bV_{k,l,p}$ in accordance with
(\ref{hommap}). Generally, this is not  automatic because
$g$ in (\ref{homm1}) is reconstructed up to exact forms
which can {\it a priori} belong to other spaces  $\bV_{n,m,p}$.
For instance, the homotopy
operators defined with respect to $Z\pm Y$ rather than
$Z$, which were  used in the early works on HS interactions
(see e.g. \cite{Vasiliev:1990bu}), do not respect
the filtration  of the spaces $\bV_{k,l,p}$ and do not fit
the above scheme. {\it A posteriori}, we realize that it
is  property (\ref{hommap}) that to large extent determines the algebraic setup
underlying the HS equations.

\subsection{Local HS algebra}

A \emph{local HS algebra} $\Spl\subset \Sp$ is defined as follows
\be
\label{Gg}
\Spl  = \oplus_{p=0}^M \Spl_p\q \Spl_p\subset \bV_{p-1,M-p-1+\epsilon,p}\q \forall
\epsilon >0
\,.
\ee
The difference between $\Spl$ and
$\Sp$ is  dominated by any rational behavior in $1-\tau$.
From {\it Lemmas 4,5} and {\it Theorem 2}  the analogues of {\it Lemma 7} and
{\it Theorems 3,4} follow\\

\noindent
{\it Lemma 7 $'$}:
$
\Spl_p *\Spl_q \subset \Spl_{p+q}\,.
$
\\

\noindent
{\it Theorem 3 $'$}: $\Spl$ is an algebra.\\

\noindent
{\it Theorem 4 $'$}: $\Spl$  is invariant under the action of
the homotopy operator $\pp$ and  $\dr_Z$.\\

{}From Eq.~(\ref{strhs}) it follows that the supertrace of  elements of $\Sp_p$
diverges as
\be
str (\Sp_p ) \sim \int_0^1 d\tau(1-\tau)^{-1-p} \ldots\,.
\ee
In particular, the supertrace of an
element of $\Sp_0$ diverges at most logarithmically. The
supertraces of elements of $\Spl_p$ with $p>0$ diverge analogously to  $\Sp_p$.
However elements of $\Spl_0$ have well-defined supertrace.
By {\it Lemma 7 $'$}  we arrive at\\

\noindent
{\it Theorem 7}:
$\Spl_0$  is an algebra endowed with the well defined supertrace for elements
(\ref{ffreg})  with
\be
\label{phipr}
\phi(W;U;\tau) = \tau^{-1} (1-\tau)^{M-1+\epsilon} \phi'(W;U;\tau)
\ee
obeying
\be
\label{strcon}
\int d^{M} U d^{M} V
\phi'(  V;U;\tau)
\exp{-i( V_A U^A)}<\infty \,.
\ee
\\

In particular, (\ref{strcon}) holds for any $\phi'(  V;U;\tau)$ polynomial in $V$ and $U$.

The algebras $\Sp_0$ and $\Spl_0$ play an important r\'ole in the analysis of \cite{funk}
where invariant functionals of the $3d$ and $4d$ HS theories are constructed as
certain projections of combinations of HS fields in the $\theta_Z$-independent
sector, that should have divergent supertrace, hence belonging to $\Sp_0/\Spl_0$.

It is also useful to introduce the {\it ultralocal algebra} $\Sp^{ult}$
\be
\label{Ggu}
\Sp^{ult}  = \oplus_{p=0}^M \Sp^{ult}_p\q \Sp^{ult}_p\subset \bV_{p-1,\infty,p}\,.
\ee
This contains the boundary subalgebra
of physical fields $\overline\Sp^{phys}\subset \Sp^{ult}_0$.

\section{Locality conjecture}
\label{locality}
Results of Section \ref{Sp} lead to a conjecture that may help to distinguish
between local and nonlocal functionals and field redefinitions in the HS theory.
Namely, consider a perturbative function
of the fields
$\phi=\W',\B \in \Sp$
\be
\label{local}
f(\phi) =f +\sum_{g,h,\ldots} (g_1* \phi*g_2 + h_1*\phi*h_2*\phi*h_3 +\ldots  )\,,
\ee
where summation is over various $f,g,h,\ldots$.
We call $f(\phi)$ \emph{local} if all $f,g,h,\ldots\in \Spl$, \emph{minimally nonlocal}
 if $f,g,h,\ldots \in \Sp$ and \emph{strongly nonlocal} otherwise.
Analogous terminology applies to
 field redefinitions $\phi \to \phi' = f(\phi)$.
Clearly, since ($\Sp$)$\Spl$ is an algebra, the composition of any two
(minimally nonlocal) local transformations (\ref{local}) is (minimally nonlocal)
local. Since $\Spl\subset \Sp$, any local transformation is minimally nonlocal.

The conjecture is that, in the HS theory, so defined {\it local} transformations
provide a proper generalization of the local transformations in Minkowski space.
We will call transformations {\it strongly local} if $f,g,h,\ldots\in \Spl$
are supported by polynomial $\phi(W;U;\tau)$ in (\ref{ffreg}). (Recall that,
by {\it Theorem 1}, such elements  form an algebra.)
Analogously, transformations (\ref{local}) will be called {\it (strongly)
ultralocal} for (polynomial) $f,g,h,\ldots\in  \Sp^{ult}$.
The most restrictive class is with polynomial
$f,g,h,\ldots\in  \overline\Sp^{phys}$.
{\it Maximally local} maps of this class are the closest analogues of the local maps in
Minkowski space,  describing  usual local field redefinitions of the
physical fields like $\go^1$ (\ref{w101}) and $C^0$ (\ref{BC}).

The rationale behind the locality conjecture is that,
as discussed in Section \ref{intro}, by virtue of the unfolded equations
 the behavior in the twistor-like variables $Z^A$ and $Y^A$ effectively encodes
the space-time derivative expansion  of the dynamical fields hidden in
$\phi$. For $f\in \bV_{k,l}$ (\ref{ffreg}) with
\be
\label{sub}
\phi(\theta_Z;
W;U;\tau) = \tau^{k} (1-\tau)^{l} \phi'(\theta_Z;
W;U;\tau)\,,
\ee
where, schematically,
\be
\label{anm}
\phi'(\theta_Z;
W;U;\tau) =\sum_{n,m} a_{nm}(\theta_Z; \tau) W^n U^m
\ee
we obtain that
\be
\label{dom}
f(\theta_Z;
Z;Y;\tau) \preceq\sum_{n,m,r} \f{(k+n+r)!(m+l)! }{(k+n+r+m+l+1)!} \bar  a_{nm}(\theta_Z)
 Z^{n+r} Y^{m+r}\,,
\ee
where $\bar  a_{nm}(\theta_Z)$ dominates $a_{nm}(\theta_Z; \tau)$ over $\tau$ and
 $\preceq$ implies that $f(\theta_Z;Z;Y;\tau)$ is dominated by the {\it r.h.s.}
 of (\ref{dom}). As anticipated, due to the exponential $\exp{i\tau Z_A Y^A}$
 in (\ref{ffreg}), the expansion of the $f(\theta_Z; Z;Y;\tau)$ is infinite even
 for polynomial $\phi'(W;U;\tau)$. The expansion coefficients
on the {\it r.h.s.} of (\ref{dom})  decrease faster for higher $k$ and $l$.

Note that solutions of differential equations always have a nonlocal form in terms of initial data with the nonlocality represented by the  Green functions.
Degree of nonlocality  depends on the type of differential equations. Inclusion
of higher derivatives makes the Green functions more nonlocal. The results of
this paper suggest that the degree of nonlocality increases
if the interaction terms contain central elements
that do not belong to $\Sp$.

The algebra $\Spl$ is  more restrictive than $\Sp$. The condition that
the composition of any two local transformations has to be local is
quite strong. From this
perspective the fact that $\Spl$ forms an algebra  is of crucial significance.

The proposed characterization of locality is  justified by the fact that the
HS equations reconstruct a solution in the class $\Sp$. This implies
that by a shift of the HS fields valued in $\Sp$ it is possible to remove the
{\it r.h.s.} of (\ref{SS}), \ie it is cohomologically trivial in $\Sp$. Such a shift is
analogous to a shift of a spin-one or spin-two field removing the current or stress tensor
from the {\it r.h.s.} of the Maxwell or Einstein equations which is essentially nonlocal
containing the Green function in the standard space-time picture. On the other hand,
the {\it r.h.s.} of (\ref{SS}) cannot be removed
by a field redefinition from  $\Spl$.

The integrating flow of \cite{Prokushkin:1998bq}, leading to a
nonlocal transformation removing the currents from the {\it r.h.s.} of the
$3d$ HS equations,
is in fact strongly nonlocal, mapping the HS fields to $\Sp^{tot}$
 as one can see from explicit formulae in  \cite{Prokushkin:1998bq,Vasiliev:1999ba}.
Hence, as anticipated, it describes a nonlocal transformation in the setup of this paper.

General field redefinitions (\ref{local}) can affect the form of the HS equations.
Gauge transformations preserve the form of the HS equations.
From the locality perspective it is important to distinguish between allowed  and
not allowed gauge transformations. The conjecture is that gauge transformations (\ref{dw})
with the gauge parameters valued in $\Sp$ are allowed.
This is necessary, in particular, to make the perturbative analysis uniquely defined,
allowing to gauge fix to zero the gauge freedom in  $\varepsilon\in \Sp$ in (\ref{homm1}).
The gauge transformations with gauge parameters beyond $\Sp$, belonging for instance
to $\Sp^{tot}$, cannot not be regarded as allowed gauge transformations since the
transformed fields do not belong to $\Sp$.

The specification of the class of allowed gauge transformation is one of the
most  significant conclusions of this paper. This is important
in particular  in the context of application of quasi gauge transformations
that ``gauge away" the space-time dependence as, e.g., in
\cite{Vasiliev:1990bu,Giombi:2012ms}. Applying this trick one has to make sure
that the final answer  belongs to $\Sp$.

A subtlety in the analysis of locality is that proper
prescription of the HS dynamics demands that  physical fields $\phi^{phys}$
(see Section \ref{sketch}) are identified  with those in
the $\dr_Z$ cohomology in all orders of the perturbation theory
\be
\phi^{phys}= \phi^{phys}_1 + \phi^{phys}_2 +\phi^{phys}_3+\ldots\,,
\ee
where $\phi_n$ describes a possible order-$n$ contribution.
In other words, for a proper
interpretation of a  solution of the HS equations with HS fields valued in $\Sp$
one has to check that their
reduction to the  $\dr_Z$ cohomology does not acquire nonlinear corrections
in higher orders of the perturbative expansion.
The {\it canonical}  evaluation of perturbative corrections
based on the application of the homotopy operator $\p^*$
automatically obeys this condition since higher-order corrections do not
  contribute to the $\dr_Z$ cohomology. Hence, the conjecture
is that a proper solution to the HS equations should be locally equivalent
to some canonical solution, \ie equivalent up to transformations
(\ref{local}) from $\Spl$ or gauge transformations from $\Sp$.

\section{Conclusion}
\label{conc}
Results  of this paper are anticipated to shed light on the long-standing
issue of locality in HS theories via the conjecture that local field redefinitions
are from the class $\Spl$ while the
gauge transformations are from the class $\Sp$. On the other hand,  field
redefinitions beyond the class $\Spl$ and  gauge transformations beyond the class
$\Sp$ should be regarded as nonlocal. In agreement with the expectation
of \cite{Prokushkin:1998bq},  this conjecture rules out the
pseudolocal field redefinitions resulting from the integrating flow of
\cite{Prokushkin:1998bq}. It would be interesting to test the locality
conjecture by its application to  other problems including
the following.

The BH  solutions of
\cite{Didenko:2009td,Iazeolla:2011cb,Bourdier:2014lya}  need further investigation
to clarify whether  they belong to the proper class of functions. A closely
related problem is that the solution of \cite{Didenko:2009td} was obtained in a
nonstandard gauge that complicates its physical interpretation. {}From the perspective of this paper the problem is to bring the available solution
to the class $\Sp$ by a formal gauge transformation.
The analysis can also be affected by the fact that  the
behavior of the BH solutions in the $Y$ variables
is essentially nonpolynomial, containing certain
 Fock vacuum factors \cite{Didenko:2009td}. This may require  further specification of
the proper functional classes in the $Y$ variables that can affect  the issue of locality.
{}From this perspective, being instrumental, consideration of this paper is not complete.
Further investigation  in spirit of, e.g., \cite{Soloviev:2013iba}
 (and references therein) would be desirable to elaborate appropriate restrictions
 on the coefficients $a_{nm}$ in (\ref{anm}) beyond the class of polynomials.

Analysis of this paper
applies to the vacuum solutions
that start with $Q$ (\ref{Q}) and to  the homotopy operator of the
form (\ref{pzp}). As such, it is fully appropriate for the twistorial
HS theories of \cite{more,Prokushkin:1998bq} but is less straightforward
for other HS models. In particular,  the construction of the
vectorial HS models in any dimension of \cite{Vasiliev:2003ev} involves
the $sp(2)$ generators $\tau_{ij}$ which restrict the field pattern to the
algebra $S$ of
$sp(2)$ singlets followed by quotiening the ideal $I$ spanned by elements
of the form $\tau_{ij}* f^{ij}$ or $f^{ij} * \tau_{ij}$. To apply the consideration
of this paper to the vectorial models of \cite{Vasiliev:2003ev} it is first of all
necessary to check whether $\tau_{ij}\in \Sp$. This is likely to be true since
$\tau_{ij}$ starts with  elements of $\Sp$ bilinear in $Y$ in the lowest order
and is reconstructed in higher orders by the equations  analogous to
the HS equations (\ref{SB}) (for more detail see
\cite{Vasiliev:2003ev,Bekaert:2005vh}). The next step is to choose representatives in $S/I$.
A proper definition should be such that $S/I\in \Sp$.
Practically, the choice of proper representatives of $S/I\in \Sp$ may
be a tricky part of  the  analysis of the vectorial $HS$ theory.

The suggested setup  helps to uncover the origin of the form of
the nonlinear HS equations as encoding certain cohomology. This interpretation
is useful for the search of more general models and approaches. For instance,
in the lowest order of the perturbative expansion, the {\it r.h.s.} of the HS equations
is valued in the boundary part of $\Sp/\Spl$, \ie
that at the boundary values of the homotopy parameter $\tau=0$ or $1$ (see Section
\ref{inbou}). This property
is  anticipated to be true for other  HS theories  including those considered
in \cite{funk}, hence clarifying their structure. A surprising output of this paper is
that certain central elements of the HS algebra, that
naively  look harmless, are ruled out from possible extensions of
the {\it r.h.s.} of (\ref{SS}) as not belonging
to the HS field algebra $\Sp$. In particular, the terms proportional to $\theta_A\theta^A$ are not allowed to
appear anywhere in the HS equations except for the first term of (\ref{SS}). This
is  related to  another unexpected conclusion  that the
$Z$-derivative  is an outer derivation of the HS  field algebra $\Sp$ because
 $Q$ (\ref{Q}) does not belong to $\Sp$. The condition that central elements that do not
 belong to $\Sp$ should not be allowed in the extended HS equations significantly restricts
 the space of extended HS systems and possible invariants among, for instance,
 those considered
  in \cite{Boulanger:2011dd,funk}.

Analysis of this paper suggests the following interesting option.
Consider the trivially looking system
\be
\label{WW}
\W*\W = \B\,,
\ee
\be
\label{WB}
 [\W\,,\B]_*=0
\ee
 invariant under the gauge transformations
\be
\label{dwgen}
\delta \W = \xi +[\W\,,\gvep]_*\,,
\ee
\be
\label{dwgen1}
 \delta \B = \{\W\,,\xi\}_* +[\B\,,\gvep]_*\,,
\ee
where the gauge parameters  $\gvep$ and $\xi$ are differential forms of even and odd
degrees, respectively. This system is empty if $\W$ can be entirely gauge fixed to zero by the $\xi$
transformation.

Suppose however that
\be
\label{gx}
\W, \B,\gvep \in \Sp\q \xi \in \Spl\,
\ee
which  assignment just fits the locality conjecture because the $\gvep$ transformation
is the HS gauge transformation while the $\xi$ transformation shifting the fields
$\W$ can in particular describe their nonlinear field redefinitions  argued
to belong to $\Spl$ in Section \ref{locality}.

Eq.~(\ref{WB}) then  implies that $\B$ is perturbatively $Q$-closed. On the other hand, the
first term in  gauge transformation (\ref{dwgen1}) washes out the $Q$-exact part of $\B$.
As a result, upon gauge fixing the $\xi$-symmetry, the remaining components of $\B$ are
in $\Sp/\Spl $. Although the remaining part of equation (\ref{WW}) looks analogous to
 HS equations (\ref{SS}), there is an essential difference
because so defined $\B$ is valued in the full cohomology $\Sp/\Spl $ rather than in the
boundary cohomology with $\tau=0$ or $1$ as the {\it r.h.s.} of (\ref{SS}).

As explained in Section \ref{inbou}, the full cohomology $\Sp/\Spl $ is represented by
functions of $2M$ variables. This space  is far larger than the boundary
cohomology represented by functions of a single variable $Y$ at $\tau=0$ or $Z$ at $\tau=1$.
This makes it tempting to speculate that  system (\ref{WW})-(\ref{gx}) contains enough auxiliary degrees of
freedom to represent the $3d$ and $4d$ off-shell HS systems.
There is a  subtlety however  that it is not obvious whether this system respects the
usual local Lorentz symmetry. We leave this problem for the future.

\label{Conclusion}

\section*{Acknowledgments}
I am grateful to Vyatcheslav Didenko, Carlo Iazeolla, Nikita Misuna,
 Mikhail Soloviev and, especially,  Olga Gelfond for  useful comments and discussions.
This research was supported by the Russian Science Foundation Grant No 14-42-00047.

\newcounter{appendix}
\setcounter{appendix}{1}
\renewcommand{\theequation}{\Alph{appendix}.\arabic{equation}}
\addtocounter{section}{1} \setcounter{equation}{0}
 \renewcommand{\thesection}{\Alph{appendix}.}

 \addtocounter{section}{1}
\addcontentsline{toc}{section}{\,\,\,\,\,\,\,Appendix A. Proof of {\it Lemma 5}}

\section*{Appendix A. Proof of {\it Lemma 5}}
Let $f(\theta_Z;Z;Y)\in V_{k,l,p}$\,.
Changing the integration variables $(s,\tau)$ to $(s,t),
t=s\tau
$
and using the substitution (\ref{sub}), (\ref{pfdz}) gives
\be
\pp f(\theta_Z;Z;Y)=\int \f{d s}{s^2}{dt} \vartheta (t)\vartheta(s-t)\vartheta(1-s)
(1-\tau)^l \tau^k
Z^A\f{\p}{\p \theta^A}  \phi'( s \theta_Z;  t Z;(1-\tau)Y;\tau)
\exp i{t Z_A Y^A}\,.
\ee
Let us focus on the case with regular dependence on $1-\tau$. (Distributions can be considered
separately.)
Since $\tau\geq t$ and  the dependence on $1-\tau$ is demanded to be regular,
all factors of $1-\tau$ are dominated  by $1-t$. For $\phi'(  \theta_Z;   W; U;\tau) $ smooth in $\tau$ this gives
\be
\pp f(\theta_Z;Z;Y)\preceq \int {d s}{dt} \vartheta (t)\vartheta(s-t)\vartheta(1-s)
(1-t)^l t^{k} s^{p-k-2}
Z^A\f{\p}{\p \theta^A} \phi^\prime_{max}( \theta_Z;  t Z;(1-t)Y)
\exp{it Z_A Y^A}\,,
\ee
where $\preceq$ implies that
the behavior on $\tau$ is dominated by the {\it r.h.s.} where
$
\phi^\prime_{max}\big( \theta_Z;W;U)
$
is some function regular in its arguments.

Integration over $s$ gives the factor of $(1-t^{p-k-1})$.
For $p > k+1$ it is dominated by $1-t$, \ie
\be
\pp f(\theta_Z;Z;Y)\preceq \int_0^1 {dt} (1-t)^{l+1}  t^{k-1}
(t Z^A)\f{\p}{\p \theta^A}  \phi^\prime_{max}( \theta_Z;  t Z;(1-t)Y)
\exp{it Z_A Y^A}\,.
\ee

For $p< k+1$, $(1-t^{p-k-1})$ is dominated by
$t^{-p+k+1}(1-t)$  yielding
\be
\pp f(\theta_Z;Z;Y) \preceq\int_0^1 {dt} (1-t)^{l+1}  t^{p-2}
(t Z^A)\f{\p}{\p \theta^A}  \phi^\prime_{max}( \theta_Z;  t Z;(1-t)Y)
\exp{it Z_A Y^A}\,.
\ee

In the case of $p=k+1$ the integration over $s$  develops
a logarithmic dependence on $t$
\be
\pp f(\theta_Z;Z;Y)\preceq \int_0^1 {dt} (1-t)^{l}  t^{k -1}\log (t)
(t Z^A)\f{\p}{\p \theta^A}  \phi^\prime_{max}( \theta_Z;  t Z;(1-t)Y)
\exp{t Z_A Y^A}\,.
\ee
Since $\log(t)$ has a simple zero at $t\to 1$ and
logarithmic singularities at $t\to 0$ do not affect indices
of $\bV_{k,l,p}$  we finally obtain {\it Lemma 5}.

\setcounter{appendix}{2}
\renewcommand{\theequation}{\Alph{appendix}.\arabic{equation}}
\addtocounter{section}{2} \setcounter{equation}{0}
 \renewcommand{\thesection}{\Alph{appendix}.}

 \addtocounter{section}{1}
\addcontentsline{toc}{section}{\,\,\,\,\,\,\,Appendix B. Local HS algebra and Weyl star product}

\section*{Appendix B. Local HS algebra and Weyl star product}

Since polynomials of oscillators can be represented both in the Weyl
(\ie totally symmetric) and in the \wick ordering prescriptions, the \wick and Weyl star products of
polynomials define the same  algebra  in different frames.
The HS algebra is realized in terms of the particular normal-ordered star product
(\ref{star2}). The  intertwining relations between the Weyl and HS orderings are
 \be \label{intwe}
f_{W}(Z;Y) = \f{1}{(2\pi)^M}\int
dSdT \exp -i S_A T^{A} f_{HS} (Z + S; Y +T)\,,
\ee
\be
\label{intwi} f_{HS}(Z;Y) = \f{1}{(2\pi)^M} \int dSdT \exp i S_A
T^{A} f_{W} (Z +S; Y +T)\,.
\ee
Indeed, it is not difficult to check   that the star product $\star$
induced by  substitution (\ref{intwi}) from  HS star product  (\ref{star2})
yields the integral version of the Weyl-Moyal star product \cite{BerezShub}
\bee
\label{starw}
&&(f_W\star g_W)(Z;Y)=\\
&&=\f{1}{(2\pi)^{2M}}
\int d UdV  \exp{[i(-U_1{}_A V_1^A + U_2{}_A V_2^A)]}\, f_W(Z+U_1;Y+U_2)
g_W(Z+V_1;Y+V_2)\nn \,.
\eee

Being equivalent for polynomials, different star products
may be  inequivalent beyond this class. This phenomenon has clear
origin. Reordering of a monomial of any degree from one
ordering prescription to another contributes to polynomials of lower degrees.
For a nonpolynomial function, containing an infinite
number of terms, the sum of the contributions to a particular (e.g.,
constant part) may diverge. Hence, beyond the class of polynomials,
different star products may require more careful definition of the functional classes.

For $f$ (\ref{ffreg}), the map (\ref{intwe}) yields
\bee
f_W (Z;Y)=&&\ls\f{1}{(2\pi)^M} \int_0^1 d\tau \int dSdT\,\exp{[-i(1-\tau ) S_A T^A+
i\f{\tau}{1-\tau}Z_A Y^A]}\nn\\&&
\phi\big( \tau S+\f{\tau}{1-\tau}Z;Y+ (1-\tau)T;\tau\big)
\,.
\eee
Changing the variables
$
(1-\tau)T = U\,
$
yields
\bee
f_W (Z;Y)=&&\ls\f{1}{(2\pi)^M} \int_0^1 d\tau (1-\tau)^{-M}
\int dSdT\,\exp{[-i S_A T^A+
i\f{\tau}{1-\tau}Z_A Y^A]}\nn\\&&
\phi\big( \tau S+\f{\tau}{1-\tau}Z;Y+ T;\tau\big)
\,.
\eee
For $\phi( V;U;\tau)$ that behave as $(1-\tau)^{M-1+\epsilon}$
at $\tau\to 1$, the measure $d\tau (1-\tau)^{\epsilon-1}$ is well defined.
However, the $Z$-dependent terms in the exponential and the argument of $\phi$
contain the factors of $(1-\tau)^{-1}$ divergent at $\tau\to 0$. As a result,
the naive map from both $\Sp$ and $\Spl$ to the Weyl star product is ill defined.
However, for $Z=0$ the map from $\Spl$ to the Weyl star product is well defined while
that from $\Sp$ is not. This is in agreement with the fact that the supertrace is
well defined in $\Spl$ but not in $\Sp$ (recall that, in accordance with
(\ref{strweyl}), the supertrace in the Weyl star product is evaluated at $Z=0$).

Note that the more careful analysis of the relation between the two
types of star product demands to take into account the equivalence relations
(\ref{equiv}) by choosing a representative in $\Spl$ obeying
(may be appropriately modified)  condition (\ref{cla}).


\begin{thebibliography}{99}
\parindent=0pt
\parskip=0pt

%\small


%\cite{Prokushkin:1998bq}
\bibitem{Prokushkin:1998bq}
  S.~F.~Prokushkin and M.~A.~Vasiliev,
  %``Higher spin gauge interactions for massive matter fields in 3-D AdS space-time,''
  Nucl.\ Phys.\ B {\bf 545} (1999) 385
  [hep-th/9806236].

%\cite{Prokushkin:1999xq}
\bibitem{Prokushkin:1999xq}
  S.~F.~Prokushkin and M.~A.~Vasiliev,
  %``Cohomology of arbitrary spin currents in AdS(3),''
  Theor.\ Math.\ Phys.\  {\bf 123} (2000) 415
   [Teor.\ Mat.\ Fiz.\  {\bf 123} (2000) 3]
   [hep-th/9907020].


\bibitem{more} M.~A.~Vasiliev, {\it Phys. Lett.}  B {\bf 285} (1992) 225.



%\cite{Vasiliev:2003ev}
\bibitem{Vasiliev:2003ev}
  M.~A.~Vasiliev,
  %``Nonlinear equations for symmetric massless higher spin fields in (A)dS(d),''
  Phys.\ Lett.\ B {\bf 567} (2003) 139
  [hep-th/0304049].


\bibitem{BerezShub}F.A.~Berezin and M.A.~Shubin, \emph{``Schr\"{o}dinger Equation''},
Moscow University Press, Moscow, 1983.

\bibitem{V3}     M.~A.~Vasiliev, {\it Fortschr. Phys.\/} {\bf 36} (1988) 33.

\bibitem{funk}
  M.~A.~Vasiliev,
  %``Invariant Functionals in Higher-Spin Theory,''
  arXiv:1504.07289 [hep-th].

\bibitem{333}
M.~A.~Vasiliev, { Nucl.Phys.} B {\bf 793} (2008) 469, {\tt
arXiv:0707.1085 [hep-th]}.


%\cite{Boulanger:2011dd}
\bibitem{Boulanger:2011dd}
  N.~Boulanger and P.~Sundell,
  %``An action principle for Vasiliev's four-dimensional higher-spin gravity,''
  J.\ Phys.\ A  {\bf 44} (2011) 495402
  [arXiv:1102.2219 [hep-th]].

%\cite{Vasiliev:1990bu}
\bibitem{Vasiliev:1990bu}
  M.~A.~Vasiliev,
  %``Algebraic aspects of the higher spin problem,''
  Phys.\ Lett.\ B {\bf 257} (1991) 111.


%\cite{Vasiliev:1999ba}
\bibitem{Vasiliev:1999ba}
  M.~A.~Vasiliev,
  %``Higher spin gauge theories: Star-product and AdS space,''
  arXiv:hep-th/9910096.


\bibitem{Pr}M.A. Vasiliev, {\it Class. Quant. Grav.} {\bf 8}, 1387 (1991).

%\cite{Giombi:2012ms}
\bibitem{Giombi:2012ms}
  S.~Giombi and X.~Yin,
  %``The Higher Spin/Vector Model Duality,''
  J.\ Phys.\ A {\bf 46} (2013) 214003
  [arXiv:1208.4036 [hep-th]].

%\cite{Didenko:2009td}
\bibitem{Didenko:2009td}
  V.~E.~Didenko and M.~A.~Vasiliev,
  %``Static BPS black hole in 4d higher-spin gauge theory,''
  Phys.\ Lett.\ B {\bf 682} (2009) 305
   [Erratum-ibid.\ B {\bf 722} (2013) 389]
  [arXiv:0906.3898 [hep-th]].

%\cite{Iazeolla:2011cb}
\bibitem{Iazeolla:2011cb}
  C.~Iazeolla and P.~Sundell,
  %``Families of exact solutions to Vasiliev's 4D equations with spherical, cylindrical and biaxial symmetry,''
  JHEP {\bf 1112} (2011) 084
  [arXiv:1107.1217 [hep-th]].

%\cite{Bourdier:2014lya}
\bibitem{Bourdier:2014lya}
  J.~Bourdier and N.~Drukker,
  %``On Classical Solutions of 4d Supersymmetric Higher Spin Theory,''
  arXiv:1411.7037 [hep-th].




%\cite{Soloviev:2013iba}
\bibitem{Soloviev:2013iba}
  M.~A.~Soloviev,
  %``Algebras with convergent star products and their representations in Hilbert spaces,''
  J.\ Math.\ Phys.\  {\bf 54} (2013) 073517
  [arXiv:1312.6571 [math-ph]].

%\cite{Bekaert:2005vh}
\bibitem{Bekaert:2005vh}
  X.~Bekaert, S.~Cnockaert, C.~Iazeolla and M.~A.~Vasiliev,
  %``Nonlinear higher spin theories in various dimensions,''
  arXiv:hep-th/0503128.
  %%CITATION = HEP-TH/0503128;%%


\end{thebibliography}
\end{document}